\documentclass[aps,prl,twocolumn,showpacs,superscriptaddress,
preprintnumbers,amsmath,amssymb]{revtex4}

\usepackage{graphicx}
\usepackage{dcolumn}
\usepackage{bm}
\usepackage{ifthen}
\usepackage{booktabs}
\usepackage{SIunits}
\usepackage{ulem}

\usepackage{comment}
\usepackage{color}

\begin{document}


\title{Tuning exciton and biexciton transition energies and fine structure splitting through hydrostatic pressure in single InGaAs quantum dots}

\author{Xuefei Wu}
\affiliation{%
State Key Laboratory for Superlattices and Microstructures,
Institute of Semiconductors, Chinese Academy of Sciences,
Beijing, 100083, People's Republic of China
}%

\author{Hai Wei}
\affiliation{%
Key Laboratory of Quantum Information, University of Science and Technology of China,
Hefei, 230026, People's Republic of China
}%

\author{Xiuming Dou}
\affiliation{%
State Key Laboratory for Superlattices and Microstructures,
Institute of Semiconductors, Chinese Academy of Sciences,
Beijing, 100083, People's Republic of China
}%

\author{Kun Ding}
\affiliation{%
State Key Laboratory for Superlattices and Microstructures,
Institute of Semiconductors, Chinese Academy of Sciences,
Beijing, 100083, People's Republic of China
}%

\author{Ying Yu}
\affiliation{%
State Key Laboratory for Superlattices and Microstructures,
Institute of Semiconductors, Chinese Academy of Sciences,
Beijing, 100083, People's Republic of China
}%

\author{Haiqiao Ni}
\affiliation{%
State Key Laboratory for Superlattices and Microstructures,
Institute of Semiconductors, Chinese Academy of Sciences,
Beijing, 100083, People's Republic of China
}%

\author{Zhichuan Niu}
\affiliation{%
State Key Laboratory for Superlattices and Microstructures,
Institute of Semiconductors, Chinese Academy of Sciences,
Beijing, 100083, People's Republic of China
}%

\author{Yang Ji}
\affiliation{%
State Key Laboratory for Superlattices and Microstructures,
Institute of Semiconductors, Chinese Academy of Sciences,
Beijing, 100083, People's Republic of China
}%

\author{Shushen Li}
\affiliation{%
State Key Laboratory for Superlattices and Microstructures,
Institute of Semiconductors, Chinese Academy of Sciences,
Beijing, 100083, People's Republic of China
}%

\author{Desheng Jiang}
\affiliation{%
State Key Laboratory for Superlattices and Microstructures,
Institute of Semiconductors, Chinese Academy of Sciences,
Beijing, 100083, People's Republic of China
}%

\author{Guang-can Guo}
\affiliation{%
Key Laboratory of Quantum Information, University of Science and Technology of China,
Hefei, 230026, People's Republic of China
}%

\author{Lixin He}
\email{helx@ustc.edu.cn}
\affiliation{%
Key Laboratory of Quantum Information, University of Science and Technology of China,
Hefei, 230026, People's Republic of China
}%

\author{Baoquan Sun}
\email{bqsun@semi.ac.cn}
\affiliation{%
State Key Laboratory for Superlattices and Microstructures,
Institute of Semiconductors, Chinese Academy of Sciences,
Beijing, 100083, People's Republic of China
}%

\date{\today}

\begin{abstract}
We demonstrate that the exciton and biexciton emission energies as well as
exciton fine structure splitting (FSS) in single (In,Ga)As/GaAs quantum dots
(QDs) can be efficiently tuned using hydrostatic pressure in situ in an
optical cryostat at up to 4.4 GPa. The maximum exciton emission energy shift
was up to 380 meV, and the FSS was up to 180 $\mu$eV.
We successfully produced a biexciton antibinding-binding transition in QDs,
which is the key experimental condition that generates color- and
polarization-indistinguishable photon pairs from the cascade of biexciton
emissions and that generates entangled photons via a time-reordering
scheme. We perform atomistic pseudopotential calculations on realistic
(In,Ga)As/GaAs QDs to understand the physical mechanism underlying the
hydrostatic pressure-induced effects.

\end{abstract}

\pacs{78.67.Hc, 07.35.+k, 78.55.Cr, 42.50.-p}
\maketitle



Self-assembled semiconductor quantum dots (QDs) have considerable potential
for use as fundamental building blocks in future quantum information
applications. However, so far it is impossible to use QD growth techniques for
precisely controlling QD properties, which is essential for such
applications. Therefore, externally tuning the QD properties post-growth is
extremely important. One the most prominent examples is polarization-entangled
photon pair emission through a biexciton (XX) cascade process in QDs, which
requires that the different polarized photons are energetically
indistinguishable. However, the underlying asymmetry for self-assembled
(In,Ga)As/GaAs QDs leads to splitting in degenerate bright exciton (X) states
(fine structure splitting, FSS), which is typically tens of
$\mu$eV~\cite{gammon96a,bayer02,gammon96b,bester03},
and much larger than the radiative linewidth ($\sim$ 1.0 $\mu$eV);
therefore, photon entanglement is destroyed~\cite{stevenson06,hafenbrak07}.
Tuning techniques, such as electric~\cite{bennett10,gerardot07,vogel07}, magnetic~\cite{hudson07}, and strain
fields~\cite{trotta12,ding10,jons11,seidl06,dou08,gong11,wang12} used to erase the FSS have been
explored.
Uniaxial and biaxial stresses have been used to tune the QD
structural symmetry, exciton binding energies and FSS.
However, the strain that can be generated using such techniques is limited to
approximately tens of MPa, which corresponds to a spectral shift by only
several meVs for QD peak emissions~\cite{ding10,jons11}.
Herein, we report high-pressure research (up to tens of GPa) for individual
QDs using the diamond anvil cell (DAC), which has been widely used to study
metal-semiconductor transitions, electronic structures, and optical
transitions in bulk crystals and
microstructures~\cite{jayaraman83,ma04,itskevich98,li94}. An exciton emission
line shift in ensemble InAs/GaAs QDs is approximately 500 meV at 8
GPa~\cite{ma04}, which is much larger than the QD peak shift induced by a
piezoelectric actuator PMN-PT~\cite{ding10}.
However, tuning the QD structural symmetry, exciton transition states and FSS
for an individual QD using the DAC has not been reported.

In this work, we demonstrate that the exciton (X) emission energy, FSS and
biexciton (XX) binding energy can be successively tuned for extremely large
ranges using hydrostatic pressure at up to 4.4 GPa. The emission energy, FSS
and XX binding energy almost increase linearly with increased pressure. The
maximum exciton emission energy shift and FSS change can extend to 380 meV and
180 $\mu$eV, respectively, which is considerably greater than through other
techniques. By tuning the applied pressure, color-indistinguishable photons
from the biexciton and exciton emission decay through a cascade, and across
generation color coincidence for biexciton and exciton transitions are
generated. Therefore, entangled photon pairs are generated via the proposed
``time reordering'' scheme~\cite{avron08}.
We also perform atomistic pseudopotential calculations on realistic
(In,Ga)As/GaAs QDs under hydrostatic stress to discern the physical mechanisms
underlying the effects induced by the hydrostatic pressure.

\begin{figure}
\centering
\includegraphics[width=2.2 in,angle=-90]{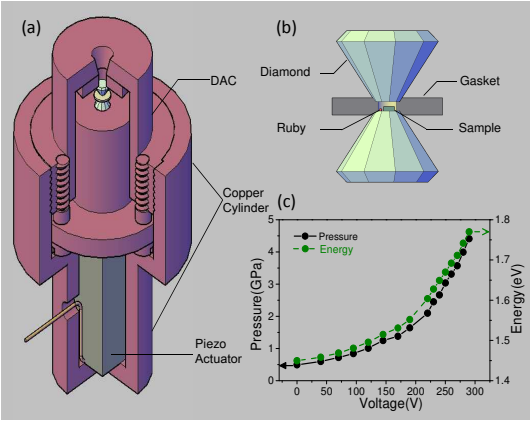}
\caption{%
(Color online) (a) Schematic drawing of the successively applied pressure
  device of the diamond anvil cell (DAC), where the DAC and piezo actuator are
  assembled together via home-made copper cylinder. (b) The DAC chamber,
  showing the positions of the QD sample and ruby in the DAC. (c) Measured
  pressure in the DAC chamber (solid black circles) and the corresponding QD
  excitonic PL peak energy (solid green circles) as a function of actuator
  voltage. At zero voltage, an initial pressure of 0.5 GPa was generated by four driven screws.
}
\label{fig:exp_equipment}
\end{figure}

The investigated (In,Ga)As/GaAs QD samples with low QD density were grown
using molecular beam epitaxy (MBE) on a semi-insulating GaAs substrate with
excitonic emission energies at 1.35-1.43 eV.
Figure~\ref{fig:exp_equipment}(a) shows the DAC pressure device used for
tuning QD photoluminescence (PL) in situ using the optical cryostat. To fit
the QD samples into the DAC chamber [indicated in
Fig.~\ref{fig:exp_equipment}(b)], the samples were mechanically thinned to a
total thickness of approximately 20 $\mu$m and then cut into  pieces
approximately 100$\times$100 $\mu$m$^2$. Condensed argon was used as the
pressure-transmitting medium in the DAC, which can be used to apply
hydrostatic pressure up to 9 GPa~\cite{jayaraman83,shimizu01}. The initial
pressure can be adjusted at room temperature by driving screws and can be
determined in situ using the ruby R1 fluorescence line shift.
To successively tune the X and XX transition energies and fine structure
splitting (FSS) by pressure at a low temperature, a novel and easily
controlled version of the DAC shown in Fig.~\ref{fig:exp_equipment}(a)
was developed by combining the well-known DAC with a piezo actuator. This
device can successively generate pressures up to several GPa for the QD
samples studied at a low temperature using PL measurements, and the maximum
applied pressure depends on the actuator stroke length. The QD sample in the
DAC is cooled to 20 K through a continuous-flow liquid helium cryostat and
excited by a He-Ne laser at the wavelength 632.8 nm. The excitation laser was
focused to a $\sim$ 2 $\mu$m spot on the sample using a microscope objective
(NA: 0.35). The PL was collected using the same objective, spectrally filtered
through a 0.5 m monochromator, and detected using a silicon-charge coupled
device (CCD). A $\lambda$/2 wave plate and linear polarizer were used to distinguish
horizontal (H) and vertical (V) linear polarization for PL components. By
carefully following the changing X and XX PL energies using the polarization
angle, we measure FSS with an $\sim$ 10 $\mu$eV accuracy by fitting the
experimental data to a sinusoidal function~\cite{ghali12}.
Figure~\ref{fig:exp_equipment}(c) displays the measured pressure values and
excitonic emission energies at 20 K as a function of actuator voltage
(Piezo-ceramics: PSt 150/10$\times$10/40). We clearly show that pressure can
be successively tuned in situ using an optical cryostat from 0.5 to 4.4 GPa
through a piezo actuator, and the corresponding blue shift for the excitonic
PL peak energy is $\sim$ 310 meV.

\begin{figure}
\centering
\includegraphics[width=2.5 in]{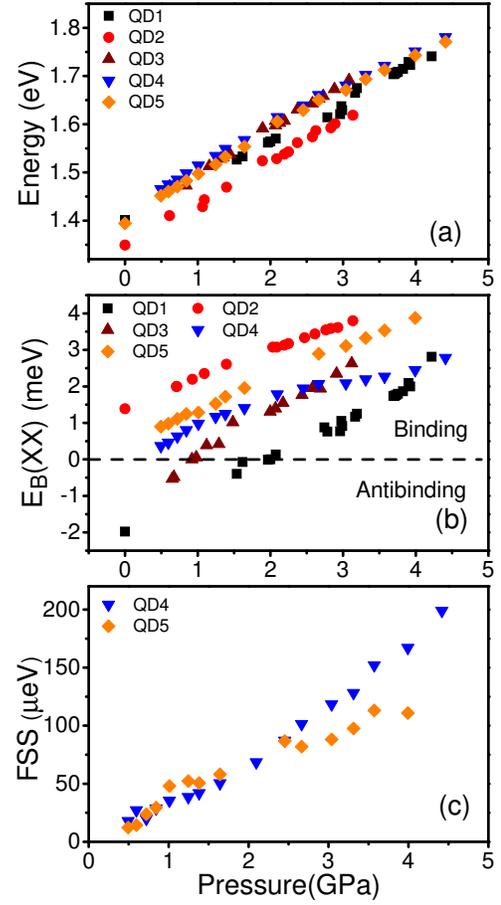}
\caption{%
(Color online) (a) Exciton emission energies as a function of pressure for
  QD1-QD5. (b) Biexciton binding energies  as a function of pressure for
  QD1-QD5, showing biexciton antibinding-binding transitions under pressure
  for QD1 and QD3. (c) FSS of QD4 and QD5 as a function of pressure.
}
\label{fig:exp_data}
\end{figure}

Figure~\ref{fig:exp_data}(a) depicts the exciton emission energies as a
function of the hydrostatic pressure from 0 to 4.4 GPa for QD1-QD5. At 0 GPa,
the QD exciton emission energies are 1.401, 1.349, 1.406, 1.432 and 1.394 eV,
respectively.
The exciton emission energies for the five QDs studied herein increased
linearly with the applied pressure. The blue shift for the QD1 peak energy at
4.22 GPa is approximately 330 meV, which is much larger than previously
reported shifts ( $\sim$ 10 meV) from uniaxial or biaxial stresses using
conventional methods~\cite{ding10,dou08,jons11,seidl06,trotta12}. The
experimental data were linearly fit, which generated pressure coefficients for
QD1-QD5 of 82, 87, 93, 81 and 85 meV/GPa, respectively; such values are
consistent with the reported pressure coefficients for ensemble quantum dot~\cite{ma04}.

Figure~\ref{fig:exp_data}(b) shows the biexciton binding energies for QD1-QD5
as a function of hydrostatic pressure at up to 4.4 GPa. The biexciton binding energy is defined as
$E_B(XX)$=$E_{X}$-$E_{XX}$, where $E_X$ and $E_{XX}$ are the X and XX emission
energies, respectively.
When $E_B(XX)>0$, the biexciton is in the ``binding'' state, wherein the two
excitons are attracted. When $E_B(XX)<0$, the biexciton is in the
``antibinding'' state, wherein the two excitons are repulsive.
For the QDs studied herein, $E_B(XX)$ increases as a function of hydrostatic
pressure up to 4.4 GPa. For QD1 and QD3, the biexcitons are in an antibinding
state at zero pressure and gradually progress to the binding state at
approximately 1 and 2 GPa, respectively (i.e., $E_B(XX)$=0), where the exciton
and biexciton are ``color-indistinguishable''.

\begin{figure}
\centering
\includegraphics[width=1.8 in,angle=-90]{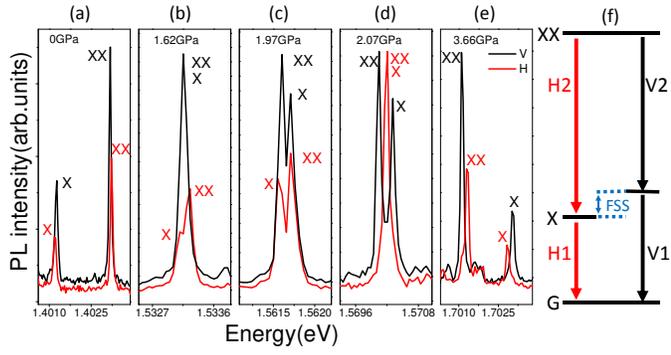}
\caption{%
(Color online) (a)-(e) Polarization-resolved PL spectra for the QD1  X and XX
  emission lines under different pressures, where red lines (black lines)
  correspond to the horizontal (H) and vertical (V) polarized photons,
  respectively. At 1.62 (b) and 2.07 (d) GPa, color-indistinguishable photon
  pairs are generated by an XX-X cascade emissions for V- and H- polarized
  photons, respectively. At 1.97 (c) GPa, across generation color coincidence
  for XX and X transition energies is achieved. (f) Level schemes showing the
  XX-X cascade at 0 GPa.
}
\label{fig:exp_polarization}
\end{figure}

To demonstrate the biexciton antibinding-binding transitions under pressure in
greater detail, we plotted the polarization-resolved PL spectra for the QD1 X
and XX emission lines under different hydrostatic pressures, as shown in
Fig.~\ref{fig:exp_polarization}(a)-(e),
wherein the red and black lines correspond to the horizontal (H) and vertical
(V) polarized photons, respectively. At zero pressure, both XX emission
energies, $E(H2)$ and $E(V2)$, are higher than the X emission energy, $E(H1)$
and $E(V1)$ [see also the scheme in Fig. \ref{fig:exp_polarization}(f)].
In addition, $E(H2)$ is slightly larger than $E(V2)$ at FSS $\sim$ 50 $\mu$eV.
Under pressure, the blue shift for the X emission energy (82 meV/GPa) is more
rapid than for XX (81 meV/GPa). Therefore, with increasing pressure, the
V-polarized XX and X emission lines first degenerate at 1.62 GPa, as shown in
Fig.~\ref{fig:exp_polarization}(b), and then the H-polarized emission lines
degenerate at 2.07 GPa, as shown in Fig.~\ref{fig:exp_polarization}(d). In
such instances, color-indistinguishable photon pairs are generated by an XX-X
cascade emission at 1.62 GPa for V-polarized photons or at 2.07 GPa for
H-polarized photons. Therefore, it is expected that the indistinguishable
two-photon streams will be produced by adjusting a time delay between the XX
and X emissions, wherein the time delay is approximately 0.4
ns~\cite{chang09}. Remarkably, at 1.97 GPa, across generation color
coincidence for XX and X transition energies was generated (i.e.,
$E(H1)$=$E(V2)$ and $E(V1)$=$E(H2)$). This is a key condition for entangled
photon generation via the proposed time reordering scheme~\cite{avron08}. When
pressure was further increased, the separation between the XX and X emission
lines again increased, as shown in
Fig.~\ref{fig:exp_polarization}(e) at 3.66 GPa.
Ding and coworkers demonstrated that biaxial strain can also tune the
biexciton binding energies~\cite{ding10}. However, because their experiment
generated a relatively small strain, biexciton antibinding-binding progression
was not observed.

FSS tuning by uniaxial strain has been studied
experimentally~\cite{seidl06,trotta12,kuklewicz12} and
theoretically\cite{singh10,gong11,wang12}, which has shown that the maximum tuned FSS value is approximately 20 $\mu$eV.
It is interesting to measure the FSS change under hydrostatic pressure.
Figure~\ref{fig:exp_data}(c) depicts the FSS for QD4 and QD5 as a function of
pressure at 20 K in the range 0.5 to 4.4 GPa. The figure clearly demonstrates
that increasing pressure produces an approximately linear increase in FSS with the slope 44 and 28
$\mu$eV/GPa for QD4 and QD5, respectively, which generates a total FSS shift
as large as $\sim$ 180 and 100 $\mu$eV for QD4 and QD5, respectively. Similar
results were observed from other investigated (In,Ga)As/GaAs QDs, which
indicates that such a large shift is typical for FSS under hydrostatic
pressure.


To understand the experimental results, we calculated the electronic and
optical properties for the In$_{1-x}$Ga$_{x}$As/GaAs QDs under hydrostatic
pressure using an atomistic empirical pseudopotential method
(EPM) ~\cite{williamson00}. The optimized QD
structures are obtained by the valence force filed method \cite{keating66}.
We then calculate the electron/hole single-particle energies and
wave functions using the linear combination of bulk bands (LCBB)
method~\cite{wang99b}. The exciton and biexciton energies are calculated
via the configuration interaction (CI) method~\cite{franceschetti99}.
Herein, we present results for three QDs: (i) a lens-shaped InAs/GaAs
QDs with the height $h$=1.5 nm and base diameter $b$=12 nm; (ii) a lens-shaped
In$_{0.8}$Ga$_{0.2}$As/GaAs QDs with $h$=1.5 nm and $b$=12, 15 nm; and (iii) In$_{0.8}$Ga$_{0.2}$As/GaAs
QDs with $h$=2.5 and the elliptical major (minor) axis $a$=10 nm
($b$=7.5 nm) along the [1$\bar{1}$0] ([110]) crystal direction.

\begin{figure}
\centering
\includegraphics[width=0.5\textwidth]{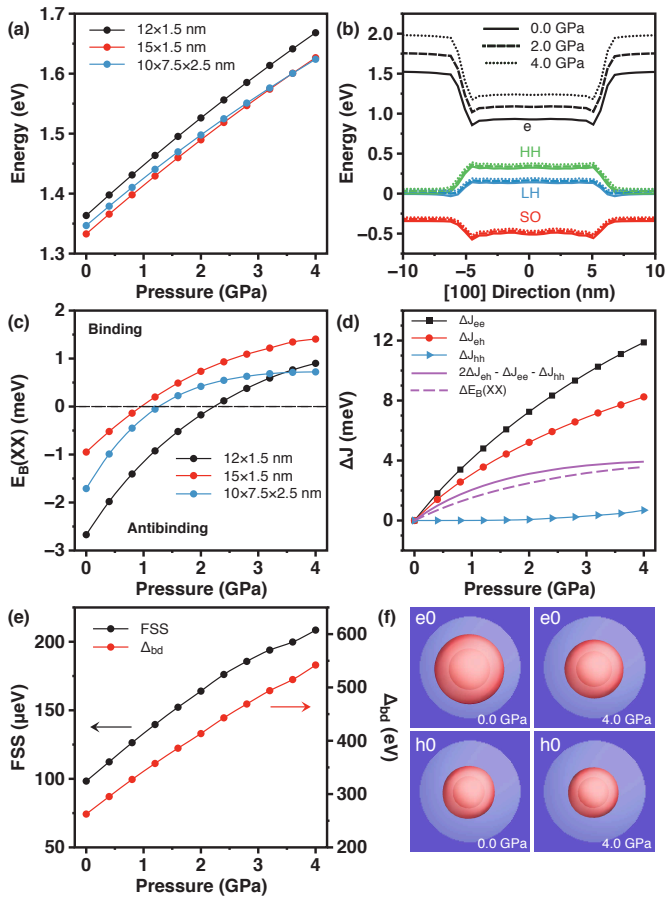}
\caption{
(Color online) The calculated results of:
(a) The exciton emission energies of (In,Ga)As/GaAs QDs as
a function of hydrostatic pressure.
(b) The band offsets of the InAs/GaAs dot under the 0.0, 2.0 and 4.0 GPa
hydrostatic pressure. The black, green, blue and red lines indicate conduction
(e), heavy-hole (HH), light-hole (LH) and
spin-orbit (SO) bands, respectively.
(c) The biexciton binding energies as a function of hydrostatic
pressure in (In,Ga)As/GaAs QDs.
(d) The changes of direct Coulomb integrals of the lowest electron and hole
states in the 12$\times$1.5 nm In$_{0.8}$Ga$_{0.2}$As/GaAs QD as a function of
the applied pressure.
(e) The FSS in the 10$\times$7.5$\times$2.5 nm In$_{0.8}$Ga$_{0.2}$As/GaAs QD
as a function of the applied pressure.
(f) The wave functions of the lowest electron (e0) and hole (h0) states under
0.0 and 4.0 GPa hydrostatic pressure in
the 12$\times$1.5 nm In$_{0.8}$Ga$_{0.2}$As/GaAs QD.
}
\label{fig:theory}
\end{figure}

The calculated exciton emission energies under pressure are shown in
Fig.~\ref{fig:theory}(a) and produce blue energy shifts at approximately 76
meV/GPa, which is consistent with the experimental values.
To understand the emission energy blue shift, we analyzed the band offsets and
confinement potentials for the QDs under pressure, which strongly depend on
the strain distribution in the dots and matrix. When hydrostatic pressure is
applied, the lattice constant for the matrix material GaAs decreases, which
effectively increases the lattice mismatch between the dot material InAs and
GaAs matrix. As a result, both the (absolute values of) isotropic and biaxial
strain inside the dots increase. The averaged isotropic strain
$I$=-0.072-0.011 $P$ and biaxial strain $\epsilon_{zz}-\epsilon_{xx}$=
0.12+0.0014 $P$, where $P$ is the applied hydrostatic pressure in GPa.
Figure~\ref{fig:theory}(b) depicts the strain-modified band offsets for the
conduction band (e), heavy hole (HH), light hole (LH) and spin-orbit (SO)
bands through the dot center under P =0, 2 and 4 GPa. Whereas the band offset
change is small for holes, the band offset changes dramatically for the
conduction band. Under pressure, the electron bands move significantly toward
the higher energy. The confinement potential also increased dramatically with
increasing pressure, which is the major reason for the observed experimental
results.

Because the electron-hole Coulomb energy change is relatively small
[see Fig.~\ref{fig:theory}(d)],
the change in exciton emission energy can be estimated using the electron-hole single-particle gap $E_g$, which can be written as follows:
\begin{equation}
E_g(\tensor{\epsilon}) = E_g(0) + a_g I + b_v(\epsilon_{zz}-\epsilon_{xx})\, ,
\end{equation}
where $\tensor{\epsilon}$ is the strain tensor inside the InAs dots,
$a_g$=$-$6.08 eV is the hydrostatic deformation potential for the band gap,
and $b_v$=$-$1.8 eV is the biaxial deformation potential for the valence band
maximum. Therefore, we estimated that the exciton PL blue shift under
hydrostatic pressure is 82 meV/GPa for pure InAs/GaAs QDs, which is consistent
with the experimental values and EPM calculations. We note that the
hydrostatic pressure is much more efficient at tuning the exciton emission
energy than uniaxial stress ($\sim$10
$\mu$eV/MPa)~\cite{jons11,kuklewicz12,wang13}.

The calculated XX binding energies $E_B(XX)$ are presented
in Fig.~\ref{fig:theory}(c).
We found that the biexciton tends toward antibinding in small QDs under zero
pressure. When the pressure increases, the $E_B(XX)$ for the dots calculated
increased. The binding energy tends to be saturated at very high pressure. The
XX binding energy for the  In$_{0.8}$Ga$_{0.2}$As/GaAs QDs
with $b$=12 nm and $h$=1.5 nm is consistent with the experimental QD1.

In the calculation, we found that it is important to include many
electron/hole energy levels for the correct XX binding energies using the CI
calculations, and the XX binding energy change as a function of pressure can
be observed only using the lowest energy conduction and valence bands (i.e.,
Hartree-Fock approximation). Such observations suggest that the XX binding energies did not change due to the correlated energies; primarily,
such changes are due to changes in the direct Coulomb integrals between the lowest electron and hole states, as follows:
\begin{equation}
\Delta E_B(XX) \approx 2\Delta J_{eh}-\Delta J_{ee}-\Delta J_{hh}\, ,
\label{eq:binding}
\end{equation}
where $J_{ee}$, $J_{hh}$ and $J_{eh}$ are the direct electron-electron, hole-hole and electron-hole Coulomb integrals, respectively.
As shown in Fig.~\ref{fig:theory}(d), whereas $J_{ee}$, and $J_{eh}$ increase rapidly with pressure, $J_{hh}$ is approximately flat.
The solid purple line describes the changing exciton binding energy calculated
using Eq.~(\ref{eq:binding}), which is consistent with the dashed purple line
from the EPM calculations. To understand how the Coulomb integrals change
under pressure, we compared the lowest electron and hole wave functions in
Fig.~\ref{fig:theory}(f) at 0.0 GPa and 4.0 GPa
for the 12$\times$1.5 nm In$_{0.8}$Ga$_{0.2}$As/GaAs QDs.
We found that, whereas the hole wave function shape primarily does not change,
the electron becomes much more localized due to the band offset changes shown
in  Fig.~\ref{fig:theory}(b),
which explains the Coulomb integral changes under pressure.

Finally, we examined FSS under hydrostatic pressure. The FSS calculated as a function of pressure is shown in
Fig.~\ref{fig:theory}(e), which increases dramatically with applied pressure
and is consistent with the experimental data in Fig.~\ref{fig:exp_data}(c). It
is surprising that FSS changes under hydrostatic pressure, which does not
change the QD symmetry. However, because the electron wave functions are more
localized under pressure, an electron-hole would have a larger effective
overlap under pressure, which increases the exchange energies (e.g., the
dark-bright splitting
$\Delta_{bd}$, which is also shown in Fig.~\ref{fig:theory}(e)).
It has been shown that FSS can be roughly estimated as $\sim
2\eta \Delta_{bd}$~\cite{wang13}, where $\eta$ is the HH-LH mixing parameter;
therefore, as $\Delta_{bd}$ increases, FSS increases, as clearly demonstrated
in Fig.~\ref{fig:theory}(e).

To summarize, we experimentally and theoretically investigated the effects of
hydrostatic pressure on the exciton and biexciton transition energies as well
as FSS in single InGaAs QDs. The excitonic emission energies and FSS can be
tuned in situ by applying hydrostatic pressure in an optical cryostat for
changes over a wide energy range. The observed exciton emission energy blue
shift and FSS change were as large as $\sim$ 380 meV and $\sim$ 180 $\mu$eV,
respectively, which is greater than the values from other strain-adjusting
techniques. Tuning the QD optical properties over such a larger spectral range
yields great advantages for future QD applications, such as for generating
color-indistinguishable photon pairs from the biexciton and exciton emission
decay cascades or generating entangled photon pairs via a time-reordering
scheme. Furthermore, it expected that photon antibunching for optical
communication band QD emission can be measured using the pressure-induced blue
shift into the spectral range detected by sufficiently powerful silicon
avalanche photodiodes.


BS and LH acknowledge support from the Chinese National Fundamental Research Program (Grant Nos. 2013CB922304,
2013CB933304, 2011CB921200, 2009CB929301), Chinese National Natural
Science Funds (Grant No. 90921015), and National Natural Science Funds for
Distinguished Young Scholars.



\begin{thebibliography}{32}
\expandafter\ifx\csname natexlab\endcsname\relax\def\natexlab#1{#1}\fi
\expandafter\ifx\csname bibnamefont\endcsname\relax
  \def\bibnamefont#1{#1}\fi
\expandafter\ifx\csname bibfnamefont\endcsname\relax
  \def\bibfnamefont#1{#1}\fi
\expandafter\ifx\csname citenamefont\endcsname\relax
  \def\citenamefont#1{#1}\fi
\expandafter\ifx\csname url\endcsname\relax
  \def\url#1{\texttt{#1}}\fi
\expandafter\ifx\csname urlprefix\endcsname\relax\def\urlprefix{URL }\fi
\providecommand{\bibinfo}[2]{#2}
\providecommand{\eprint}[2][]{\url{#2}}

\bibitem[{\citenamefont{Gammon et~al.}(1996{\natexlab{a}})\citenamefont{Gammon,
  Snow, Shanabrook, Katzer, and Park}}]{gammon96a}
\bibinfo{author}{\bibfnamefont{D.}~\bibnamefont{Gammon}},
  \bibinfo{author}{\bibfnamefont{E.~S.} \bibnamefont{Snow}},
  \bibinfo{author}{\bibfnamefont{B.~V.} \bibnamefont{Shanabrook}},
  \bibinfo{author}{\bibfnamefont{D.~S.} \bibnamefont{Katzer}},
  \bibnamefont{and} \bibinfo{author}{\bibfnamefont{D.}~\bibnamefont{Park}},
  \bibinfo{journal}{Phys. Rev. Lett.} \textbf{\bibinfo{volume}{76}},
  \bibinfo{pages}{3005} (\bibinfo{year}{1996}{\natexlab{a}}).

\bibitem[{\citenamefont{Bayer et~al.}(2002)\citenamefont{Bayer, Ortner, Stern,
  Kuther, Gorbunov, Forchel, Hawrylak, Fafard, Hinzer, Reinecke
  et~al.}}]{bayer02}
\bibinfo{author}{\bibfnamefont{M.}~\bibnamefont{Bayer}},
  \bibinfo{author}{\bibfnamefont{G.}~\bibnamefont{Ortner}},
  \bibinfo{author}{\bibfnamefont{O.}~\bibnamefont{Stern}},
  \bibinfo{author}{\bibfnamefont{A.}~\bibnamefont{Kuther}},
  \bibinfo{author}{\bibfnamefont{A.~A.} \bibnamefont{Gorbunov}},
  \bibinfo{author}{\bibfnamefont{A.}~\bibnamefont{Forchel}},
  \bibinfo{author}{\bibfnamefont{P.}~\bibnamefont{Hawrylak}},
  \bibinfo{author}{\bibfnamefont{S.}~\bibnamefont{Fafard}},
  \bibinfo{author}{\bibfnamefont{K.}~\bibnamefont{Hinzer}},
  \bibinfo{author}{\bibfnamefont{T.~L.} \bibnamefont{Reinecke}},
  \bibnamefont{et~al.}, \bibinfo{journal}{Phys. Rev. B}
  \textbf{\bibinfo{volume}{65}}, \bibinfo{pages}{195315}
  (\bibinfo{year}{2002}).

\bibitem[{\citenamefont{Gammon et~al.}(1996{\natexlab{b}})\citenamefont{Gammon,
  Snow, Shanabrook, Katzer, and Park}}]{gammon96b}
\bibinfo{author}{\bibfnamefont{D.}~\bibnamefont{Gammon}},
  \bibinfo{author}{\bibfnamefont{E.~S.} \bibnamefont{Snow}},
  \bibinfo{author}{\bibfnamefont{B.~V.} \bibnamefont{Shanabrook}},
  \bibinfo{author}{\bibfnamefont{D.~S.} \bibnamefont{Katzer}},
  \bibnamefont{and} \bibinfo{author}{\bibfnamefont{D.}~\bibnamefont{Park}},
  \bibinfo{journal}{Science} \textbf{\bibinfo{volume}{5}}, \bibinfo{pages}{87}
  (\bibinfo{year}{1996}{\natexlab{b}}).

\bibitem[{\citenamefont{Bester et~al.}(2003)\citenamefont{Bester, Nair, and
  Zunger}}]{bester03}
\bibinfo{author}{\bibfnamefont{G.}~\bibnamefont{Bester}},
  \bibinfo{author}{\bibfnamefont{S.}~\bibnamefont{Nair}}, \bibnamefont{and}
  \bibinfo{author}{\bibfnamefont{A.}~\bibnamefont{Zunger}},
  \bibinfo{journal}{Phys. Rev. B} \textbf{\bibinfo{volume}{67}},
  \bibinfo{pages}{161306} (\bibinfo{year}{2003}).

\bibitem[{\citenamefont{Stevenson et~al.}(2006)\citenamefont{Stevenson, Young,
  Atkinson, Cooper, Ritchie, and Shields}}]{stevenson06}
\bibinfo{author}{\bibfnamefont{R.~M.} \bibnamefont{Stevenson}},
  \bibinfo{author}{\bibfnamefont{R.~J.} \bibnamefont{Young}},
  \bibinfo{author}{\bibfnamefont{P.}~\bibnamefont{Atkinson}},
  \bibinfo{author}{\bibfnamefont{K.}~\bibnamefont{Cooper}},
  \bibinfo{author}{\bibfnamefont{D.~A.} \bibnamefont{Ritchie}},
  \bibnamefont{and} \bibinfo{author}{\bibfnamefont{A.~J.}
  \bibnamefont{Shields}}, \bibinfo{journal}{Nature}
  \textbf{\bibinfo{volume}{439}}, \bibinfo{pages}{179} (\bibinfo{year}{2006}).

\bibitem[{\citenamefont{Hafenbrak et~al.}(2007)\citenamefont{Hafenbrak, Ulrich,
  Michler, Wang, Rastelli, and Schmidt}}]{hafenbrak07}
\bibinfo{author}{\bibfnamefont{R.}~\bibnamefont{Hafenbrak}},
  \bibinfo{author}{\bibfnamefont{S.~M.} \bibnamefont{Ulrich}},
  \bibinfo{author}{\bibfnamefont{P.}~\bibnamefont{Michler}},
  \bibinfo{author}{\bibfnamefont{L.}~\bibnamefont{Wang}},
  \bibinfo{author}{\bibfnamefont{A.}~\bibnamefont{Rastelli}}, \bibnamefont{and}
  \bibinfo{author}{\bibfnamefont{O.~G.} \bibnamefont{Schmidt}},
  \bibinfo{journal}{New Journal of Physics} \textbf{\bibinfo{volume}{9}},
  \bibinfo{pages}{315} (\bibinfo{year}{2007}).

\bibitem[{\citenamefont{Bennett et~al.}(2010)\citenamefont{Bennett, Pooley,
  Stevenson, Ward, Patel, Boyer de~la Giroday, Sk\"{o}ld, Farrer, Nicoll,
  Ritchie et~al.}}]{bennett10}
\bibinfo{author}{\bibfnamefont{A.~J.} \bibnamefont{Bennett}},
  \bibinfo{author}{\bibfnamefont{M.~A.} \bibnamefont{Pooley}},
  \bibinfo{author}{\bibfnamefont{R.~M.} \bibnamefont{Stevenson}},
  \bibinfo{author}{\bibfnamefont{M.~B.} \bibnamefont{Ward}},
  \bibinfo{author}{\bibfnamefont{R.~B.} \bibnamefont{Patel}},
  \bibinfo{author}{\bibfnamefont{A.}~\bibnamefont{Boyer de~la Giroday}},
  \bibinfo{author}{\bibfnamefont{N.}~\bibnamefont{Sk\"{o}ld}},
  \bibinfo{author}{\bibfnamefont{I.}~\bibnamefont{Farrer}},
  \bibinfo{author}{\bibfnamefont{C.~A.} \bibnamefont{Nicoll}},
  \bibinfo{author}{\bibfnamefont{D.~A.} \bibnamefont{Ritchie}},
  \bibnamefont{et~al.}, \bibinfo{journal}{Nature Phys.}
  \textbf{\bibinfo{volume}{6}}, \bibinfo{pages}{947} (\bibinfo{year}{2010}).

\bibitem[{\citenamefont{Gerardot et~al.}(2007)\citenamefont{Gerardot, Seidl,
  Dalgarno, Warburton, Granados, Garcia, Kowalik, Krebs, Karrai, Badolato
  et~al.}}]{gerardot07}
\bibinfo{author}{\bibfnamefont{B.~D.} \bibnamefont{Gerardot}},
  \bibinfo{author}{\bibfnamefont{S.}~\bibnamefont{Seidl}},
  \bibinfo{author}{\bibfnamefont{P.~A.} \bibnamefont{Dalgarno}},
  \bibinfo{author}{\bibfnamefont{R.~J.} \bibnamefont{Warburton}},
  \bibinfo{author}{\bibfnamefont{D.}~\bibnamefont{Granados}},
  \bibinfo{author}{\bibfnamefont{J.~M.} \bibnamefont{Garcia}},
  \bibinfo{author}{\bibfnamefont{K.}~\bibnamefont{Kowalik}},
  \bibinfo{author}{\bibfnamefont{O.}~\bibnamefont{Krebs}},
  \bibinfo{author}{\bibfnamefont{K.}~\bibnamefont{Karrai}},
  \bibinfo{author}{\bibfnamefont{A.}~\bibnamefont{Badolato}},
  \bibnamefont{et~al.}, \bibinfo{journal}{Appl. Phys. Lett.}
  \textbf{\bibinfo{volume}{90}}, \bibinfo{pages}{041101}
  (\bibinfo{year}{2007}).

\bibitem[{\citenamefont{Vogel et~al.}(2007)\citenamefont{Vogel, Ulrich,
  Hafenbrak, Michler, Wang, Rastelli, and Schmidt}}]{vogel07}
\bibinfo{author}{\bibfnamefont{M.~M.} \bibnamefont{Vogel}},
  \bibinfo{author}{\bibfnamefont{S.~M.} \bibnamefont{Ulrich}},
  \bibinfo{author}{\bibfnamefont{R.}~\bibnamefont{Hafenbrak}},
  \bibinfo{author}{\bibfnamefont{P.}~\bibnamefont{Michler}},
  \bibinfo{author}{\bibfnamefont{L.}~\bibnamefont{Wang}},
  \bibinfo{author}{\bibfnamefont{A.}~\bibnamefont{Rastelli}}, \bibnamefont{and}
  \bibinfo{author}{\bibfnamefont{O.~G.} \bibnamefont{Schmidt}},
  \bibinfo{journal}{Appl. Phys. Lett.} \textbf{\bibinfo{volume}{91}},
  \bibinfo{pages}{051904} (\bibinfo{year}{2007}).

\bibitem[{\citenamefont{Hudson et~al.}(2007)\citenamefont{Hudson, Stevenson,
  Bennett, Young, Nicoll, Atkinson, Cooper, Ritchie, and Shields}}]{hudson07}
\bibinfo{author}{\bibfnamefont{A.~J.} \bibnamefont{Hudson}},
  \bibinfo{author}{\bibfnamefont{R.~M.} \bibnamefont{Stevenson}},
  \bibinfo{author}{\bibfnamefont{A.~J.} \bibnamefont{Bennett}},
  \bibinfo{author}{\bibfnamefont{R.~J.} \bibnamefont{Young}},
  \bibinfo{author}{\bibfnamefont{C.~A.} \bibnamefont{Nicoll}},
  \bibinfo{author}{\bibfnamefont{P.}~\bibnamefont{Atkinson}},
  \bibinfo{author}{\bibfnamefont{K.}~\bibnamefont{Cooper}},
  \bibinfo{author}{\bibfnamefont{D.~A.} \bibnamefont{Ritchie}},
  \bibnamefont{and} \bibinfo{author}{\bibfnamefont{A.~J.}
  \bibnamefont{Shields}}, \bibinfo{journal}{Phys. Rev. Lett.}
  \textbf{\bibinfo{volume}{99}}, \bibinfo{pages}{266802}
  (\bibinfo{year}{2007}).

\bibitem[{\citenamefont{Trotta et~al.}(2012)\citenamefont{Trotta, Zallo, Ortix,
  Atkinson, Plumhof, van~den Brink, Rastelli, and Schmidt}}]{trotta12}
\bibinfo{author}{\bibfnamefont{R.}~\bibnamefont{Trotta}},
  \bibinfo{author}{\bibfnamefont{E.}~\bibnamefont{Zallo}},
  \bibinfo{author}{\bibfnamefont{C.}~\bibnamefont{Ortix}},
  \bibinfo{author}{\bibfnamefont{P.}~\bibnamefont{Atkinson}},
  \bibinfo{author}{\bibfnamefont{J.~D.} \bibnamefont{Plumhof}},
  \bibinfo{author}{\bibfnamefont{J.}~\bibnamefont{van~den Brink}},
  \bibinfo{author}{\bibfnamefont{A.}~\bibnamefont{Rastelli}}, \bibnamefont{and}
  \bibinfo{author}{\bibfnamefont{O.~G.} \bibnamefont{Schmidt}},
  \bibinfo{journal}{Phys. Rev. Lett.} \textbf{\bibinfo{volume}{109}},
  \bibinfo{pages}{147401} (\bibinfo{year}{2012}).

\bibitem[{\citenamefont{Ding et~al.}(2010)\citenamefont{Ding, Singh, Plumhof,
  Zander, K\ifmmode~\check{r}\else \v{r}\fi{}\'apek, Chen, Benyoucef, Zwiller,
  D\"orr, Bester et~al.}}]{ding10}
\bibinfo{author}{\bibfnamefont{F.}~\bibnamefont{Ding}},
  \bibinfo{author}{\bibfnamefont{R.}~\bibnamefont{Singh}},
  \bibinfo{author}{\bibfnamefont{J.~D.} \bibnamefont{Plumhof}},
  \bibinfo{author}{\bibfnamefont{T.}~\bibnamefont{Zander}},
  \bibinfo{author}{\bibfnamefont{V.}~\bibnamefont{K\ifmmode~\check{r}\else
  \v{r}\fi{}\'apek}}, \bibinfo{author}{\bibfnamefont{Y.~H.}
  \bibnamefont{Chen}},
  \bibinfo{author}{\bibfnamefont{M.}~\bibnamefont{Benyoucef}},
  \bibinfo{author}{\bibfnamefont{V.}~\bibnamefont{Zwiller}},
  \bibinfo{author}{\bibfnamefont{K.}~\bibnamefont{D\"orr}},
  \bibinfo{author}{\bibfnamefont{G.}~\bibnamefont{Bester}},
  \bibnamefont{et~al.}, \bibinfo{journal}{Phys. Rev. Lett.}
  \textbf{\bibinfo{volume}{104}}, \bibinfo{pages}{067405}
  (\bibinfo{year}{2010}).

\bibitem[{\citenamefont{J\"ons et~al.}(2011)\citenamefont{J\"ons, Hafenbrak,
  Singh, Ding, Plumhof, Rastelli, Schmidt, Bester, and Michler}}]{jons11}
\bibinfo{author}{\bibfnamefont{K.~D.} \bibnamefont{J\"ons}},
  \bibinfo{author}{\bibfnamefont{R.}~\bibnamefont{Hafenbrak}},
  \bibinfo{author}{\bibfnamefont{R.}~\bibnamefont{Singh}},
  \bibinfo{author}{\bibfnamefont{F.}~\bibnamefont{Ding}},
  \bibinfo{author}{\bibfnamefont{J.~D.} \bibnamefont{Plumhof}},
  \bibinfo{author}{\bibfnamefont{A.}~\bibnamefont{Rastelli}},
  \bibinfo{author}{\bibfnamefont{O.~G.} \bibnamefont{Schmidt}},
  \bibinfo{author}{\bibfnamefont{G.}~\bibnamefont{Bester}}, \bibnamefont{and}
  \bibinfo{author}{\bibfnamefont{P.}~\bibnamefont{Michler}},
  \bibinfo{journal}{Phys. Rev. Lett.} \textbf{\bibinfo{volume}{107}},
  \bibinfo{pages}{217402} (\bibinfo{year}{2011}).

\bibitem[{\citenamefont{Seidl et~al.}(2006)\citenamefont{Seidl, Kroner,
  H\"{o}gele, Karrai, Warburton, Badolato, and Petroff}}]{seidl06}
\bibinfo{author}{\bibfnamefont{S.}~\bibnamefont{Seidl}},
  \bibinfo{author}{\bibfnamefont{M.}~\bibnamefont{Kroner}},
  \bibinfo{author}{\bibfnamefont{A.}~\bibnamefont{H\"{o}gele}},
  \bibinfo{author}{\bibfnamefont{K.}~\bibnamefont{Karrai}},
  \bibinfo{author}{\bibfnamefont{R.~J.} \bibnamefont{Warburton}},
  \bibinfo{author}{\bibfnamefont{A.}~\bibnamefont{Badolato}}, \bibnamefont{and}
  \bibinfo{author}{\bibfnamefont{P.~M.} \bibnamefont{Petroff}},
  \bibinfo{journal}{Appl. Phys. Lett.} \textbf{\bibinfo{volume}{88}},
  \bibinfo{pages}{203113} (\bibinfo{year}{2006}).

\bibitem[{\citenamefont{Dou et~al.}(2008)\citenamefont{Dou, Sun, Wang, Ma,
  Zhou, Huang, Ni, and Niu}}]{dou08}
\bibinfo{author}{\bibfnamefont{X.~M.} \bibnamefont{Dou}},
  \bibinfo{author}{\bibfnamefont{B.~Q.} \bibnamefont{Sun}},
  \bibinfo{author}{\bibfnamefont{B.~R.} \bibnamefont{Wang}},
  \bibinfo{author}{\bibfnamefont{S.~S.} \bibnamefont{Ma}},
  \bibinfo{author}{\bibfnamefont{R.}~\bibnamefont{Zhou}},
  \bibinfo{author}{\bibfnamefont{S.~S.} \bibnamefont{Huang}},
  \bibinfo{author}{\bibfnamefont{H.~Q.} \bibnamefont{Ni}}, \bibnamefont{and}
  \bibinfo{author}{\bibfnamefont{Z.~C.} \bibnamefont{Niu}},
  \bibinfo{journal}{Chin. Phys. Lett.} \textbf{\bibinfo{volume}{25}},
  \bibinfo{pages}{1120} (\bibinfo{year}{2008}).

\bibitem[{\citenamefont{Gong et~al.}(2011)\citenamefont{Gong, Zhang, Guo, and
  He}}]{gong11}
\bibinfo{author}{\bibfnamefont{M.}~\bibnamefont{Gong}},
  \bibinfo{author}{\bibfnamefont{W.}~\bibnamefont{Zhang}},
  \bibinfo{author}{\bibfnamefont{G.-C.} \bibnamefont{Guo}}, \bibnamefont{and}
  \bibinfo{author}{\bibfnamefont{L.}~\bibnamefont{He}},
  \bibinfo{journal}{Physical Review Letters} \textbf{\bibinfo{volume}{106}},
  \bibinfo{pages}{227401} (\bibinfo{year}{2011}).

\bibitem[{\citenamefont{Wang et~al.}(2012)\citenamefont{Wang, Gong, Guo, and
  He}}]{wang12}
\bibinfo{author}{\bibfnamefont{J.}~\bibnamefont{Wang}},
  \bibinfo{author}{\bibfnamefont{M.}~\bibnamefont{Gong}},
  \bibinfo{author}{\bibfnamefont{G.-C.} \bibnamefont{Guo}}, \bibnamefont{and}
  \bibinfo{author}{\bibfnamefont{L.}~\bibnamefont{He}},
  \bibinfo{journal}{Applied Physics Letters} \textbf{\bibinfo{volume}{101}},
  \bibinfo{pages}{063114} (\bibinfo{year}{2012}).

\bibitem[{\citenamefont{Jayaraman}(1983)}]{jayaraman83}
\bibinfo{author}{\bibfnamefont{A.}~\bibnamefont{Jayaraman}},
  \bibinfo{journal}{Rev. Mod. Phys.} \textbf{\bibinfo{volume}{55}},
  \bibinfo{pages}{65} (\bibinfo{year}{1983}).

\bibitem[{\citenamefont{Ma et~al.}(2004)\citenamefont{Ma, Wang, Su, Fang, Ding,
  Niu, and Li}}]{ma04}
\bibinfo{author}{\bibfnamefont{B.~S.} \bibnamefont{Ma}},
  \bibinfo{author}{\bibfnamefont{X.~D.} \bibnamefont{Wang}},
  \bibinfo{author}{\bibfnamefont{F.~H.} \bibnamefont{Su}},
  \bibinfo{author}{\bibfnamefont{Z.~L.} \bibnamefont{Fang}},
  \bibinfo{author}{\bibfnamefont{K.}~\bibnamefont{Ding}},
  \bibinfo{author}{\bibfnamefont{Z.~C.} \bibnamefont{Niu}}, \bibnamefont{and}
  \bibinfo{author}{\bibfnamefont{G.~H.} \bibnamefont{Li}}, \bibinfo{journal}{J.
  Appl. Phys.} \textbf{\bibinfo{volume}{95}}, \bibinfo{pages}{933}
  (\bibinfo{year}{2004}).

\bibitem[{\citenamefont{Itskevich et~al.}(1998)\citenamefont{Itskevich, Lyapin,
  Troyan, Klipstein, Eaves, Main, and Henini}}]{itskevich98}
\bibinfo{author}{\bibfnamefont{I.~E.} \bibnamefont{Itskevich}},
  \bibinfo{author}{\bibfnamefont{S.~G.} \bibnamefont{Lyapin}},
  \bibinfo{author}{\bibfnamefont{I.~A.} \bibnamefont{Troyan}},
  \bibinfo{author}{\bibfnamefont{P.~C.} \bibnamefont{Klipstein}},
  \bibinfo{author}{\bibfnamefont{L.}~\bibnamefont{Eaves}},
  \bibinfo{author}{\bibfnamefont{P.~C.} \bibnamefont{Main}}, \bibnamefont{and}
  \bibinfo{author}{\bibfnamefont{M.}~\bibnamefont{Henini}},
  \bibinfo{journal}{Phys. Rev. B} \textbf{\bibinfo{volume}{58}},
  \bibinfo{pages}{R4250} (\bibinfo{year}{1998}).

\bibitem[{\citenamefont{Li et~al.}(1994)\citenamefont{Li, Go\~ni, Syassen,
  Brandt, and Ploog}}]{li94}
\bibinfo{author}{\bibfnamefont{G.~H.} \bibnamefont{Li}},
  \bibinfo{author}{\bibfnamefont{A.~R.} \bibnamefont{Go\~ni}},
  \bibinfo{author}{\bibfnamefont{K.}~\bibnamefont{Syassen}},
  \bibinfo{author}{\bibfnamefont{O.}~\bibnamefont{Brandt}}, \bibnamefont{and}
  \bibinfo{author}{\bibfnamefont{K.}~\bibnamefont{Ploog}},
  \bibinfo{journal}{Phys. Rev. B} \textbf{\bibinfo{volume}{50}},
  \bibinfo{pages}{18420} (\bibinfo{year}{1994}).

\bibitem[{\citenamefont{Avron et~al.}(2008)\citenamefont{Avron, Bisker,
  Gershoni, Lindner, Meirom, and Warburton}}]{avron08}
\bibinfo{author}{\bibfnamefont{J.~E.} \bibnamefont{Avron}},
  \bibinfo{author}{\bibfnamefont{G.}~\bibnamefont{Bisker}},
  \bibinfo{author}{\bibfnamefont{D.}~\bibnamefont{Gershoni}},
  \bibinfo{author}{\bibfnamefont{N.~H.} \bibnamefont{Lindner}},
  \bibinfo{author}{\bibfnamefont{E.~A.} \bibnamefont{Meirom}},
  \bibnamefont{and} \bibinfo{author}{\bibfnamefont{R.~J.}
  \bibnamefont{Warburton}}, \bibinfo{journal}{Phys. Rev. Lett.}
  \textbf{\bibinfo{volume}{100}}, \bibinfo{pages}{120501}
  (\bibinfo{year}{2008}).

\bibitem[{\citenamefont{Shimizu et~al.}(2001)\citenamefont{Shimizu, Tashiro,
  Kume, and Sasaki}}]{shimizu01}
\bibinfo{author}{\bibfnamefont{H.}~\bibnamefont{Shimizu}},
  \bibinfo{author}{\bibfnamefont{H.}~\bibnamefont{Tashiro}},
  \bibinfo{author}{\bibfnamefont{T.}~\bibnamefont{Kume}}, \bibnamefont{and}
  \bibinfo{author}{\bibfnamefont{S.}~\bibnamefont{Sasaki}},
  \bibinfo{journal}{Phys. Rev. Lett.} \textbf{\bibinfo{volume}{86}},
  \bibinfo{pages}{4568} (\bibinfo{year}{2001}).

\bibitem[{\citenamefont{Ghali et~al.}(2012)\citenamefont{Ghali, Ohtani, Ohno,
  and Ohno}}]{ghali12}
\bibinfo{author}{\bibfnamefont{M.}~\bibnamefont{Ghali}},
  \bibinfo{author}{\bibfnamefont{K.}~\bibnamefont{Ohtani}},
  \bibinfo{author}{\bibfnamefont{Y.}~\bibnamefont{Ohno}}, \bibnamefont{and}
  \bibinfo{author}{\bibfnamefont{H.}~\bibnamefont{Ohno}},
  \bibinfo{journal}{Nat. Commun.} \textbf{\bibinfo{volume}{3}},
  \bibinfo{pages}{661} (\bibinfo{year}{2012}).

\bibitem[{\citenamefont{Chang et~al.}(2009)\citenamefont{Chang, Dou, Sun,
  Xiong, Niu, Ni, and Jiang}}]{chang09}
\bibinfo{author}{\bibfnamefont{X.~Y.} \bibnamefont{Chang}},
  \bibinfo{author}{\bibfnamefont{X.~M.} \bibnamefont{Dou}},
  \bibinfo{author}{\bibfnamefont{B.~Q.} \bibnamefont{Sun}},
  \bibinfo{author}{\bibfnamefont{Y.~H.} \bibnamefont{Xiong}},
  \bibinfo{author}{\bibfnamefont{Z.~C.} \bibnamefont{Niu}},
  \bibinfo{author}{\bibfnamefont{H.~Q.} \bibnamefont{Ni}}, \bibnamefont{and}
  \bibinfo{author}{\bibfnamefont{D.~S.} \bibnamefont{Jiang}},
  \bibinfo{journal}{J. Appl. Phys.} \textbf{\bibinfo{volume}{106}},
  \bibinfo{pages}{103716} (\bibinfo{year}{2009}).

\bibitem[{\citenamefont{Kuklewicz et~al.}(2012)\citenamefont{Kuklewicz, Malein,
  Petroff, and Gerardot}}]{kuklewicz12}
\bibinfo{author}{\bibfnamefont{C.~E.} \bibnamefont{Kuklewicz}},
  \bibinfo{author}{\bibfnamefont{R.~N.~E.} \bibnamefont{Malein}},
  \bibinfo{author}{\bibfnamefont{P.~M.} \bibnamefont{Petroff}},
  \bibnamefont{and} \bibinfo{author}{\bibfnamefont{B.~D.}
  \bibnamefont{Gerardot}}, \bibinfo{journal}{Nano Lett.}
  \textbf{\bibinfo{volume}{12}}, \bibinfo{pages}{3761} (\bibinfo{year}{2012}).

\bibitem[{\citenamefont{Singh and Bester}(2010)}]{singh10}
\bibinfo{author}{\bibfnamefont{R.}~\bibnamefont{Singh}} \bibnamefont{and}
  \bibinfo{author}{\bibfnamefont{G.}~\bibnamefont{Bester}},
  \bibinfo{journal}{Phys. Rev. Lett.} \textbf{\bibinfo{volume}{104}},
  \bibinfo{pages}{196803} (\bibinfo{year}{2010}).

\bibitem[{\citenamefont{Williamson et~al.}(2000)\citenamefont{Williamson, Wang,
  and Zunger}}]{williamson00}
\bibinfo{author}{\bibfnamefont{A.~J.} \bibnamefont{Williamson}},
  \bibinfo{author}{\bibfnamefont{L.-W.} \bibnamefont{Wang}}, \bibnamefont{and}
  \bibinfo{author}{\bibfnamefont{A.}~\bibnamefont{Zunger}},
  \bibinfo{journal}{Phys.\ Rev.\ B} \textbf{\bibinfo{volume}{62}},
  \bibinfo{pages}{12963} (\bibinfo{year}{2000}).

\bibitem[{\citenamefont{Keating}(1966)}]{keating66}
\bibinfo{author}{\bibfnamefont{P.~N.} \bibnamefont{Keating}},
  \bibinfo{journal}{Phys. Rev} \textbf{\bibinfo{volume}{145}},
  \bibinfo{pages}{637} (\bibinfo{year}{1966}).

\bibitem[{\citenamefont{Wang and Zunger}(1999)}]{wang99b}
\bibinfo{author}{\bibfnamefont{L.-W.} \bibnamefont{Wang}} \bibnamefont{and}
  \bibinfo{author}{\bibfnamefont{A.}~\bibnamefont{Zunger}},
  \bibinfo{journal}{Phys.\ Rev.\ B} \textbf{\bibinfo{volume}{59}},
  \bibinfo{pages}{15806} (\bibinfo{year}{1999}).

\bibitem[{\citenamefont{Franceschetti et~al.}(1999)\citenamefont{Franceschetti,
  Fu, Wang, and Zunger}}]{franceschetti99}
\bibinfo{author}{\bibfnamefont{A.}~\bibnamefont{Franceschetti}},
  \bibinfo{author}{\bibfnamefont{H.}~\bibnamefont{Fu}},
  \bibinfo{author}{\bibfnamefont{L.-W.} \bibnamefont{Wang}}, \bibnamefont{and}
  \bibinfo{author}{\bibfnamefont{A.}~\bibnamefont{Zunger}},
  \bibinfo{journal}{Phys.\ Rev.\ B} \textbf{\bibinfo{volume}{60}},
  \bibinfo{pages}{1819} (\bibinfo{year}{1999}).

\bibitem[{\citenamefont{Wang et~al.}(unpublished)\citenamefont{Wang, Guo, and
  He}}]{wang13}
\bibinfo{author}{\bibfnamefont{J.}~\bibnamefont{Wang}},
  \bibinfo{author}{\bibfnamefont{G.-C.} \bibnamefont{Guo}}, \bibnamefont{and}
  \bibinfo{author}{\bibfnamefont{L.}~\bibnamefont{He}}
  (\bibinfo{year}{unpublished}).

\end{thebibliography}


\clearpage

\clearpage

\end{document}